\documentstyle[epsf,twoside,fleqn,psfig,espcrc2]{article}%
\begin{document}                                                        
\renewcommand{\refname}{\normalsize\bf References}
\title{%
Thermoelectric Field Effects in Low Dimensional Structure Solar Cells
}

\author{%
         Stefan Kettemann%
        \address{I. Institut f.
               Theoretische
               Physik\\ 
               Jungiusstr. 9,  
               20355
               Hamburg
               Germany 
                 },%
        \,and
        Jean-Francois Guillemoles%
        \address{ENSCP 
              Rue Pierre et Marie Curie, 75231 Paris
  France
          }%
}
%
%
\begin{abstract}
\hrule
\mbox{}\\[-0.2cm]

\noindent{\bf Abstract}\\
 Taking into account the 
 temperature gradients
 in solar cells, 
it is shown 
 that their
efficiency  can be increased beyond the Shockley- Queisser-
limit \cite{shockley}. The
 driving force for this gain is the temperature gradient between this
region and
its surroundings. 
A quantitative theory is given. Though the effect is found 
to be weak
in conventional  solar cells, it is argued that it can  be substantially
increased by
proper choice of materials and design of the  device.
 In particular, it is shown that the insertion of a 
quantum well can enhance the efficiency beyond the one of a single gap cell, 
  due to the presence of  temperature jumps at the heterojunctions. 

\hrule
\end{abstract}
\maketitle
\section{Introduction}   
  The suggestion of  Barnham and Duggan, 
   that  the efficiency
 of  solar cells can       
 be increased when low dimensional structures
 like multi quantum wells  are inserted between 
 their p-n contact \cite{barnham}, sparked off a renewed interest 
 in the derivation   of the 
   efficiency limits of ideal   solar cells.
 It was soon  realized, 
 that  detailed balance theory would result in  the limiting efficiency 
 of the lower band gap material. Thus, no enhancement beyond the 
 SQ limit is found
 (as obtained for a single band gap material, taking into account 
 radiative recombination \cite{shockley}), 
 if the  assumption of full thermalization of 
 excited electron- hole pairs in two bands is    made\cite{araujo}. 
  Recently, it has been argued   that 
 a solar cell with multibands may  exceed this efficiency limit, 
since  the assumption of fast thermalization may only be valid within each
  subband, due to slow interband nonradiative recombination.
 Thus,   under illumination 
 the charge carriers
 can have   separate quasi- Fermi- levels in each subband\cite{us,lm}, 
 which results in efficiency limits corresponding to the one of 
 multiple tandem cells\cite{henry}, even exceeding earlier estimates
 \cite{corkish}. 
  Recent experiments on multi-quamtum well solar cells seem 
  not  to be in accord with a model of two  quasi-Fermi levels\cite{jenny}.  
 It has been suggested that  low dimensional structures like 
 superlattices of quantum dots could be preferable
 for a realization of  multiband solar cells
\cite{us,honsberg}.  In these structures interband electron-phonon 
 relaxation can be reduced substantially\cite{bottleneck}, 
 although this effect can be overturned  by many-body effects\cite{efros}
 and high temperature. 
 Intersubband absorption of photons \cite{lm},
  interband transitions due to electron- electron scattering, 
 or the inverse Auger effect \cite{honsberg} can  be ways to 
 have multiple quasi- Fermi levels under detailed balance conditions, 
 und thus an increase of efficiency beyond the single gap limit.  

 Here,  another mechanism is explored, 
   which  could lead to an enhancement of 
  the efficiency of solar cells, 
 even when the local 
 distribution of charge carriers is well described by two 
 quasi-Fermi levels:  the heating of the optically active region of the 
 semiconductor. 
 As a result, the 
 thermoelectric force 
 due to temperature gradients in the  cell may enhance the output voltage. 
 There may also be an enhancement 
 of current and output voltage 
 due to  thermionic emission from   the optically 
 active region, if  a semiconductor with 
 a lower band gap than the emitter and basis is inserted, there. 
 The efficiency of semiconductor thermoelements which turn temperature
 gradients into electrical current is limited by the
 figure of merit $ T Z  = T S^2 \sigma/\kappa q^2$, with the  
 thermopower $S$, 
  the conductivity $\sigma$,  and thermal conductivity $\kappa$ \cite{ioffe},
\cite{tritt}. 
   It might be of interest to combine solar cells 
 with  thermoelements to reach a larger overall efficiency\cite{parrott}. 
  Here, we will explore if  temperature gradients
 inside a  solar cell can increase their efficiency beyond the 
 SQ limit. 
 We will assume fast thermalization of the charge carriers with the 
 lattice, neglecting hot electron effects, considered in Refs.
\cite{ross,fardi}.

 In the second chapter, the   
 limiting efficiency as function of  band gap and temperature is  
  reviewed  for a single band gap solar cell.
  In the third chapter, the  effect of 
 temperature gradients on the performance of a single band gap solar cell 
 is  explored. 
    In the final chapter, it is studied, if the insertion of a lower band gap 
 material  in the presence of the temperature gradients, 
 may enhance the efficiency further. 

 It is assumed that the surrounding of the 
 solar cell is not heated and stays at
 the temperature $T =  300 K$. 
 Furthermore, in 
 order to study the  efficiency limit, 
 we assume that   radiative recombination 
 between the conduction and valence band is predominant, 
 and nonradiative recombination can be  disregarded\cite{shockley}.
 The temperature gradient is assumed to be  caused by the
fast thermalisation
of the carriers in the optically active region. 

\section{Temperature sensitivity of Single Gap Cell}
 While a uniform heating  of the solar cell is commonly known to be 
 detrimental 
 to its efficiency, 
 there are known   exceptions,
 like GaAs heterojunction solar cells\cite{alferov}, 
 in which an  
 increase in efficiency  by  1 per cent
 was measured, when raising the   
 temperature to 350 K\cite{hovel}. 
  Let us first review therefore the 
 temperature sensitivity of a uniform  single gap solar cell. 
 From   the decrase of Carnot efficiency, $\Delta \eta = -\Delta T/T_s$
 with increasing
 temperature $T$,  one expects the trend to decreasing efficiency. 
  However, the energy gap is known to decrease with increasing $T$ 
 in most semiconductors. 
 Thus,  one could also expect an increase of efficiency when the 
 energy gap at room temperature is above the one giving optimal 
 efficiency according to SQ\cite{shockley}.  
  That is,  for energy gaps 
 above   $ E_G = 1.4 eV$.
 In order  to find out,
 which factor is dominant, the increase of photocurrent 
 due to decreased band gap, or, the increase of radiative recombination, 
 let us reconsider  the ideal efficiency as function of  temperature $T$.

The current is ideally 
 given by the charge times the rate of absorbed photons
 with energy $ E =  h \nu $, exceeding the energy gap $ E_G$, 
  subtracted by the 
 photons emitted by radiative recombination: 
\begin{equation} \label{current}
I ( V ) = I_{Solar} ( E_G )+  I_{T} ( E_G ) - I_{T} ( E_G, V  ),  
\end{equation} 
 where $q V = \epsilon_{C} - \epsilon_{V}$, 
 with the absolute value of the  electron charge, $ q$.  
   It is assumed
  that the electron distribution in the conduction and valence band 
 under illumination, is well described by a Fermi distribution 
 with a Fermi level $\epsilon_{C}$ in the conduction band and 
  $\epsilon_{V}$ in the valence band, respectively.  
  The current terms are given by 
 $  I_{Solar} ( E_G ) = - A_s q k_c N_{Solar} ( E_G)$, 
 with the cell area $ A_s$ , the solar concentration factor
 $1 < k_c < 46050$. The rate of 
 solar photons  is modeled by a  Planck distribution
 with temperature $ T_S = 5800 K $ : 
 $N_{Solar} ( E_G) = ( f_s/  4 \pi ) (c/4) 
 \int_{E_G/h}^{\infty} d \nu
 2 D_{\nu}/(\exp ( h \nu/(k_{\rm B} T_s) ) -1 )$, where 
 $  D_{\nu} = 4 \pi \nu^2/c^3$, $c$ being the velocity of light, 
 and the solar solid angle $ f_S = .000272$.
 Thus,  $  I_{Solar} ( E_G )/A_s = - 425 k_c  ( \int_{E_G/k_B T_s}^{\infty} dx
 x^2/( e^x - 1 )) A/m^2$, where $ k_B T_s = .5 eV$.  
 The  current  due to absorption of thermal photons is 
 $I_{T} ( E_G )= - A_s q  N_{T} ( E_G )$ 
with a photon rate arising from    the Planck distribution
 with  surrounding temperature $T$,  
 $N_{T} ( E_G) =  (c/2) \int_{E_G/h}^{\infty} d \nu
 2 D_{\nu}/(\exp ( h \nu/(k_{\rm B} T) ) -1 )$. It  is 
 exponentially small for $ E_G \gg k_B T$. 
 The current, lost due to radiative recombination is,  
$  I_{T} ( E_G, V  )=  - A_s q  N_{T} ( E_G, V)$, 
 where $N_{T}  ( E, V) $ is the  photon rate emitted by a 
   semiconductor with quasi Fermi- level separation $q V $
 and temperature $T$ as given by the modified Planck distribution\cite{wuerfel}, 
  $N_{T} ( E, V) =  c/2 \int_{E_G/h}^{\infty} d \nu
 2 D_{\nu}/(\exp ( ( h \nu - q V)/(k_{\rm B} T) ) -1 )$. 
 Thus, the dark current is 
  $  I_{T} ( E_G, V  )/A_s = -  435.2 ( T/ 300 K)^3 ( \int_{E_G/ k_B T
 }^{\infty} d x / ((\exp (  x - q V/(k_{\rm B} T) ) -1 )) A/m^2$. 
 The efficiency of a one band solar cell is obtained
 by optimizing  $ I(V) V/ A P_s$ with respect to the voltage V,
 giving  for  
$ E_G - q V > k_{\rm B} T$,   
\begin{equation} \label{singleeff}
\eta = \eta_0 \frac{f^2}{1+f},
\end{equation}
 where 
$\eta_0 = \frac{425}{1400} \frac{k_B T }{e V} \int_{2 E_G/e V}^{\infty} dx 
 x^2/(e^x -1)$.
 Here, $ f ( g ) = q V_m/k_B T$, is a function of 
 $g = I_{Solar}/I_T ( E_G, V=0) $, only, as  given by the 
 nonlinear equation
$ f = \ln g - \ln ( 1+f )$. 
 The decrease of the energy gap as function of temperature $T$
 is close to room  temperature, $ T = 300 K$, linear in $T$: 
 $ E_{\rm G} ( T ) =  E_{\rm G} ( 0 K ) - z_{\rm G} k_{\rm B}  T$, 
 where   $z_{\rm G}$  ranges between  
 3  for Si and 5  
 for $TiO_2$\cite{seeger}. 
 The    efficiency is found to decrease  linearly in temperature, for 
 $   E_G < 2.88 k_B T_S = 1.4 eV$.
 For
 energy gaps exceeding the maximum of the solar radiation  distribution, 
 $ E_G > 1.4 eV $, the decrease of the 
 energy gap  with increasing  temperature $T$ 
 results in an exponential enhancement of the 
 photocurrent, since the gap is in the exponentially 
 decaying part of the Planck distribution. This 
 can overcome  the 
 exponential increase of the dark current.
  Indeed, this  results in a slight increase of
 efficiency $\Delta \eta \sim .01 \% \Delta T/  K$  with temperature
 increase $\Delta T$, 
  for   $ E_G > 1.4 eV$. 
  This increase can be enhanced   by concentration $k_c$, 
 as   $ \ln k_c$.
 At higher temperatures it reaches a maximum
 at a temperature $T_M ( E_G ) $ for 
 $1.4 eV <  E_G < 3.5 eV$  and 
 decreases than strongly.
  It reaches  a maximum increase above the SQ limit 
 by  $1 \% $ at a temperature of $ 1000 K$, for 
 a band gap of $ E_G = 3.5 eV$. 
 At $ E_{\rm G}  > 3.5 eV $, the efficiency drops exponentially below 
 $ 5 \%$, and accordingly its   increase
  with temperature becomes exponentially small, although the relative 
 efficiency increase becomes larger. 
  From Eq.  (\ref{singleeff}),   for $ E_G \gg k_B T_S$ and $ T \ll T_S$ 
  the efficiency increase is 
 for an  ideal solar cell  given by 
\begin{equation}
\Delta \eta = 2.1  z_{\rm G} 10^{-2} (\frac{E_G}{eV})^3 \exp ( - 2 E_G(0)/eV)
\frac{\Delta T}{K} \%.
\end{equation}
  \section{Thermoelectric field enhancement of Single Gap Cell efficiency }
 In this section we will address the question, 
  if the  inhomogenous     heating of the solar cell under solar irradiation   yields an 
 enhancement of the output voltage and thereby an increase of 
 the efficiency of an ideal 
  gap solar cell beyond the SQ limit due to thermoelectric effects. 
  The  heating of the semiconductor
 due to the thermalization of the electron-hole pairs  in the optically
  active region produces a temperature gradient, which acts as a
 driving force on the charge carriers away from the heat source. 
  The efficiency 
 of a  thermoelectrical element alone,  is 
$\eta_{TE} = W/Q$, where $ W = I V$ is the extracted electrical 
work and $ Q = K \Delta T + S I T_i/q $ the transfered heat
 from the source of  temperature 
 $ T_i > T$ 
 with thermal conductance $K$. 
 $Q$ is smaller   
 when a current $I$ is flowing, due to the Peltier cooling, as included,
 above.
 The  efficiency limit of such a thermo element is known to be 
 given by \cite{ioffe,tritt}
 \begin{eqnarray} \label{ioffe}
\eta_{TE} = \frac{T_i -T}{T_i} \frac{( 1 + T Z   )^{1/2}}{T/T_i + ( 1 + T Z  )^{1/2}}.
\end{eqnarray}
 with  the figure of merit,  $ T Z = T S^2 \sigma /\kappa q^2$, 
 which is   generally smaller than 1\cite{tritt}.
 As a function of  the temperature gradients 
 in a  solar cell under irradiation, one can find 
 with  Eq. (\ref{ioffe}) the efficiency to  turn them  into 
 electrical work.

 However, we are rather interested 
 if the total efficiency of the solar cell, $\eta = I V/A P_s$, 
 can exceed the limit derived in the previous section, 
 when the temperature gradients are taken into account. 
   To this end, the current voltage characteristics of the solar
 cell have to be rederived.  
  The separation of the charge carriers is due to drift in  the 
 intrinsic electrical field of the p-n contact,  a small 
  ohmic driving field  
  and due to diffusion.  Due to the  temperature gradient, there is
 the additional thermoelectrical current, driving electrons to the n-doped
 basis, 
 and holes to the p-doped emitter.   
  The electron current $I$ per area $A_s$ 
 arriving at the back contact 
  through the n- doped region is thus,  
  in a classical drift diffusion model in the presence
 of a 
 quasi- Fermi level gradient $ \nabla \epsilon_c$, and the 
 temperature gradient
 $ \nabla T$, given
 by\cite{seeger} 
\begin{equation} \label{in}
I_n/A_s  =   \mu_n n_c (  \nabla  \epsilon_c - \nabla E_c 
 -  S_n \nabla T - S_{D n}  \nabla T )  , 
\end{equation}
with the electron 
 mobility  $\mu_n = q \tau/m_c $,  
 the scattering rate 
 $\tau$, and effective mass $m_c$.
 The  electron density
 is  $n_c = N_C ( T) \exp ( ( q \phi - E_c +
 \epsilon_c)/k_{\rm B} T )$,
where, for parabolic energy dispersion,  $ N_C = 2.5 (10^{19}/cm^3)
 ( m_c/m)^{3/2} ( T/300 K )^{3/2} $. 
  $S_n$ is the  thermopower of the n-doped semiconductor
 under nondegenerate doping conditions,  given by 
\begin{equation}
 S_n =    k_{\rm B}  (  \ln n_c/N_c -  r - 5/2  ) ).  
\end{equation} 
  The current, Eq. (\ref{in} ), 
  is a sum of 1.    the drift- diffusion current,
 governed by the conductivity, 
 which is in the n-doped region given by 
$ \sigma_n = q n_c  \mu_n$,  and  includes  the 
 additional term  due to  the gradient in conduction band edge  
 $\nabla E_c$. 
 2. the thermoelectric electron current, 
 where $r$ is the power of the scattering time as function of energy, 
$ \tau \sim E^r $, where
   $ r= - 1/2$ when electron scattering is dominated by  acoustic
 phonon scattering, and 
 3. the last term, 
  the electron current due to the phonon flux away from the 
 higher temperature $T_i$ region, dragging electrons with them. 
 Accordingly, the phonon drag  contribution to the thermopoer, 
 $S_D$,  
 is due to electon- phonon scattering. It can usually be disregarded
 at room temperature, since it is proportional to the 
 phonon mean free path ($S_D \ll k_{\rm B}$)\cite{herring}.  
  The hole  current arriving at the top  contact 
  through the p- doped region is accordingly given by 
$
I_p/A_s  =    \mu_p p_v (  -  \nabla \epsilon_v + \nabla E_v
 -   S_p \nabla T + S_{D p} \nabla T)    $.  
 The hole conductivity is in the p-doped region given by 
  $ \sigma_p = p_v q mu_p$,  with the  hole  mobility  $\mu_p$.
 and  the hole  density   
 $p_v = P_V ( T) \exp (- ( q \phi - E_v +
 \epsilon_v)/k_{\rm B} T )$,
 where $ P_V ( T) =  2.5 (10^{19}/cm^3)
 ( m_v/m)^{3/2} ( T/300 K )^{3/2} $. 
The thermopower of the holes in the p- doped region is given by 
$ S_p = -  k_{\rm B}  
 (  \ln p_v/P_v    -   r - 5/2  )$.

 Now, one can obtain the current voltage characteristics of the 
   solar cell, by the condition that all the current 
 created in the i-region, $ I (V_i) $ arrives at the contacts, 
 or $ I_p =   I (V_i) = I_n$, where only the voltage 
 drop across the  optical  region $V_i$ enters, since 
 radiative recombination mainly occurs there.   
  The spatial dependence of the energy gap due to the 
 temperature gradient 
 results in a gradient of the conduction band edge 
$ \nabla E_c = -   z_{\rm C} k_{\rm B} \nabla T $,  
and valence band edge, $ \nabla E_V =   z_{\rm V} k_{\rm B} \nabla T $,  
 where    $ z_{\rm C} + z_{\rm V}  = z_G $. 
 Thus,  
one obtains  for the gradient of the quasi- Fermi level in the n-doped 
basis: 
\begin{equation} 
 \nabla \epsilon_c =     (- z_{\rm C} k_{\rm B}
 + S_n) \nabla T   +  \frac{ I ( V_i)  }{
 \mu_n N_D A_s}  ,
\end{equation} 
  and a corresponding expression for the 
 gradient of the quasi- Fermi level in the emitter. 
 Note that $- \nabla  \epsilon_c/ k_{\rm B} \nabla T > 1$ is therefore  
 possible. 
  The temperature gradient is obtained from the heat flow balance, 
\begin{equation}
( 1- \eta )  P_s  = - \kappa \nabla T + T \frac{S}{q} \frac{I}{A_s} + P_O,  
\end{equation}
 where  $\eta$ is the solar cell efficiency and the 
 second term is the Peltier heat which tends to cool the 
 hot absorber for $ I < 0$,  thus reducing the temperature gradient. 
 $P_O = R I^2$ is the ohmic heat due to  the cell resistance $R$. 
 Thus, the cell efficiency  is obtained from Eqs. (1), (7), (8),   
 as $\eta = I (V_i) ( V_i + d_n \nabla \epsilon_c( V_i)/q 
 + d_p \nabla \epsilon_v (V_i)/q)/A P_s$, 
 where $d_n$, $d_p$ are the thickness of the basis and emitter, respectively, 
 and $\nabla \epsilon_v (V_i)$ is the gradient in the valence band of the p-
 doped  emitter, as given by the equivalent of Eq. (7) for the holes.
 Rather than optimizing this efficiency
 with respect to the voltage $V_i$, 
 let us  consider here first  the magnitude of the thermoelectric gain. 

 As an example,  the 
 mobility of electrons in bulk GaAs,
 is  at room temperature   $ \mu_n = 5000 cm^2/ V s$, 
 at  doping  $ N_D \approx   N_C = 5 ~~ 10^{17}/cm^3$.
With the heat conductivity, which is for GaAs about 
 $ \kappa = 45 W/K m$ and the solar 
 power  $ P_S = k_c 1400 W/m^2$, one gets  
 from $ (1 -\eta )  P_s =  \kappa  \nabla  T$ 
temperature gradients   of $ \nabla  T < 
 30 k_c K/m$ within the semiconductor solar cell,
  if all solar power is absorbed in the optically active region, only.
 Thus the temperature drop 
 from the optically active region to its basis 
 in a  solar cell of basis  thickness $ d_n = 100 \mu m$, 
 is  obtained to be $ \Delta T  =  ( 1 -\eta ) 3 k_c mK $. 
 With $ S_n \approx - z_D k_{\rm B} $, 
 where $ z_D $  is at strong doping $ z_D \approx 2$ 
 we obtain 
 from the first term of Eq. (7) 
  the  postitive thermoelectric voltage 
 $\delta V_{TE}  = 1.3  k_c \mu V $.
 The  negative  ohmic  voltage is, using  the short crircuit
 current  given above,  for GaAs,  with $ E_G = 1.4 eV$, 
 $ I_{SC}/A_s \sim 400 k_c A/m^2$ obtained 
 to be  $ \delta V_{Ohm} = - 1.0 k_c \mu 
V $. 
 Thus, 
  the change in 
 output voltage obtained from $\nabla \epsilon_c/q$, 
 Eq. (7),   is   positive, the thermoelectric effect 
 is 
 dominating the ohmic one. But,  the gain is even under strong concentration
 $ k_c = 1000$ on the order of $mV$, only.
  Can this  be improved? 
Besides strong concentration there is another way to enhance the
 effect. When the solar irradiation is parallel to the 
p-n- contact\cite{piotr},
 and the illumination is concentrated on the intrinsic region, 
 only, 
 the temperature gradients are enlarged by 
 a factor $ d_i/d$, where $ d $ is the thickness of the cell.  

 Recently, the reduction of perpendicular thermal conductivity
  of AlGaAs 
 by insertion of GaAs quantum wells by a factor $1/10$ 
 has been reported\cite{chen}, 
 due to back scattering of phonons at the heterojunction. 
  The figure of merit has been reported to be enhanced in GaAs/AlAs
 superlattices by a factor of 50 at strong doping\cite{koga}. 
  This is another motivation to study  the 
 effect of insertion of 
 quantum wells on  the perfomance of  solar cells in the next chapter. 

\section{Quantum well solar cells }

 In this section, we study if  
  the insertion  of a lower band  gap material 
 in the optical region, as proposed in Ref. [2], 
 can enhance the efficency of the solar cell beyond  the SQ limit, 
 when  temperature gradients are taken into account. 
  The transport of charge carriers out of the
 lower band gap  region is due to 
 drift in the internal electrical field, diffusion,  
    and  thermionic emission into the larger gap basis and emitter.  
 Therefore, instead of Eq. (5), valid for a homogenous cell, only, 
   the current across the hetero junction   is  given by, 
\begin{eqnarray}
I &=& - q n_L v_L \exp ( ( E_{C L } -   E_{C R })/( k_{\rm B} T_L ) )
\nonumber \\ 
 &+& 
  q n_R v_R + q n \mu \phi', 
\end{eqnarray}
 where $ n_s, v_s, E_{C s}, T_s, s = L, R $ 
 are conduction electron density, thermal velocity, conduction band edge, 
 and temperature, to the right and left of the heterostructure, 
 respectively, see Fig. 1. 
 The electrical field, $ \phi' = ( V_{np} - V_i )/ d_i $ is assumed to be 
 constant across the i-region, where $ d_i $  is   
 the thickness of the intrinsic region which is taken here to coincide
 with the thickness of the quantum well, for simplicity. 
\begin{figure}\label{fig1}   
\centerline{\epsfbox{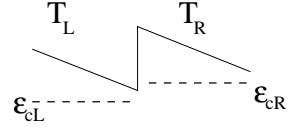}}   
\caption{ The energy conduction band edge 
 $ E_c - q \phi (x) $  with hetertojunction is sketched by the full line, 
 Quasi- Fermi levels, $\epsilon_L,\epsilon_R $
 to the right (R) and left ( L )  of the junction, 
  by broken lines. 
 }  
\end{figure}  
  Thus, all photons are assumed to be absorbed in the quantum well region, 
 giving   $ I = I ( V_i )$ as given by Eq. (\ref{current} ).    
  The average thermal velocity is  $ v_s  =(  k_{\rm B} T_s/m_s )^{1/2},
 s = L, R$, 
 and  the  densities $n_s, s = L, R $ are functions of the local temperature 
 and quani-Fermi levels as given in the 
 previous section. 
  By expansion in the  temperature
 drop $ \Delta T = T_L - T_R$ across  the heterojunction,
 a jump in the quasi- Fermi level is obtained, 
$ \Delta \epsilon_c = \epsilon_R- \epsilon_L$
given by 
\begin{equation} \label{step}
 \Delta \epsilon_c = k_{\rm B}
  \ln ( 1 + \frac{m_R}{m_L}  \frac{I}{ q v_R n_R} ) 
- S_L \Delta T,  
\end{equation}
 where the thermopower is found to be given by $ S_L = k_{\rm B} 
 (q \phi_L + \epsilon_L - E_{C R})/T$.  
 As an example let us consider a GaAs 
  quantum well between 
 an $ {\rm Al_{1/3}Ga_{2/3}As}$  emitter and basis
 with direct band gap $ E_G = 1.8 eV$\cite{mitin}, effective mass
 $ m_e = .095 m_0$,  
 and  conduction band offset, $ \Delta E_C = .68  \Delta E_{\rm G}
 = .3 eV$ \cite{koga}. 
  From the heat conductivity of an  AlAs/GaAs  quantum well 
  at room temperature of $ \kappa = 10 W/mK$, and the result of simulations
 which show that the drop in heat conductivity is mainly due to scattering of
  phonons 
 at the heterostructure boundary\cite{chen}, one can estimate 
 a temperature drop by $ \Delta T = k_c 10^{-4} K $ at the quantum well
  boundary under solar illumination with concentration $k_c$. 
 Thus, Eq. ( \ref{step} ), yields at forward voltage  a 
 jump of the quasi- Fermi level of 
$ \Delta \epsilon_c = -3 k_c 10^{-7}  (   eV - q .5 V_i )  $, which 
even   under strong concentration does not exceed $ meV$, but adds to the 
 voltage acroos the basis and emitter, as  
 derived in the previous section. 

 Thus,  the efficiency of an 
 AlGaAs/GaAs, quantum well cell is with a two band quasi-Fermi level
 distribution,  slightly enhanced beyond the SQ limit, 
 when the temperature gradients
 due to the absorption of the solar radiation in
 the  optically active i- region is taken into account, 
 and can exceed the respective efficiency of the  single gap cell.
 The thermoelectric gain is found to be small however, so that 
 it seems not to  be  possible to approach the tandem efficiency limit which
  is   for a ${\rm GaAs/Al_xGa_{1-x}As}$  tandem system, 
 without concentration, $ k_c = 1$,  for $x= 1/3$ $\eta = 38 \%$. 
\section{Discussion}
 In summary, it has been shown that the thermoelectrical effect 
due to temperature gradients in solar cells enhances their
 limiting  efficiency
 beyond the SQ limit.  
  The insertion of a quantum well into the optically active region is 
 found to enhance the thermoelectrical effect further, due to temperature 
 drops at the hetero junctions and resulting positive 
 jumps in the quasi- Fermi
 levels. This may explain   recent experiments 
 on quantum well solar cells showing  
  positive jumps in the quasi- Fermi
 levels\cite{jenny}. 
 It seems worthwhile  to extend the analyisis 
 to other heterostructure cells,  like quantum dot solar cells,  
 considering  their  reduced thermal conductivity
and favourable  themoelectrical field effects\cite{tritt}.
 Other  semiconductor heterostructures 
 than AlGaAs,  like Si/Ge have  been shown
 to be favourable thermoelectric materials\cite{tritt}.   
  Very low thermal conductivities of $ \kappa \approx 1  W/mK$, have been 
 reported in nanocrystalline $TiO_2$\cite{lee},
 so that  the thermoelectric voltage can be  a
 relevant mechanism to enhance the output voltage in 
 dye sensitized $TiO_2$ solar cells.

\end{document}